\documentclass[twocolumn,floatfix,prl]{revtex4}
\usepackage{graphicx}

\begin{document}

\title[a]{Mapping multiple photonic qubits into and out of one solid-state atomic ensemble}

\author{Imam Usmani}
\author{Mikael Afzelius}
\email{mikael.afzelius@unige.ch}
\author{Hugues de Riedmatten}
\author{Nicolas Gisin}
\affiliation{Group of Applied Physics, University of Geneva, CH-1211 Geneva 4, Switzerland}%

\date{\today}

\begin{abstract}
The future challenge of quantum communication are scalable quantum
networks, which require coherent and reversible mapping of photonic qubits onto stationary
atomic systems (quantum memories). A crucial
requirement for realistic networks is the ability to efficiently
store multiple qubits in one quantum memory.
Here we demonstrate coherent and reversible mapping of 64 optical modes
at the single photon level in the time domain onto one solid-state
ensemble of rare-earth ions. Our light-matter interface is based
on a high-bandwidth (100 MHz) atomic frequency comb, with a
pre-determined storage time $\gtrsim$ 1 $\mu$s. We can then encode
many qubits in short $<$10 ns temporal modes (time-bin qubits). We
show the good coherence of the mapping by simultaneously storing
and analyzing multiple time-bin qubits.
\end{abstract}

\maketitle

Quantum communication \cite{Gisin2007a} offers the possibility of secure transmission of
messages using quantum key distribution (QKD) \cite{Gisin2002} and teleportation of unknown quantum states \cite{Bennett1993}. Quantum communication relies on creation, manipulation and transmission of qubits in photonic channels. Photons have proven to be robust carriers of quantum information. Yet, the transmission of photons through a fiber link, for instance, is inherently a lossy process. This leads to a probabilistic nature of the outcome of experiments. In large-scale quantum networks \cite{Kimble2008} the possibility of synchronizing independent and probabilistic quantum channels will be required for scalability \cite{Duan2001,Sangouard2009a}. A quantum memory enables this by momentarily holding a photon and then releasing it when another part of the network is ready. In order to reach reasonable rates in a realistic network it will be necessary to use multiplexing \cite{Simon2007}, which requires quantum memories capable of storing many single photons in different modes.

A quantum memory requires a coherent medium with strong coupling
to a light mode. Strong and coherent interactions can been found
in ensembles of atoms \cite{Hammerer2008}, for instance alkali
atoms or rare-earth (RE) ions doped into crystals. The latter are
attractive for quantum storage applications, as they provide
solid-state systems with a large number of stationary atoms with
excellent coherence properties. Optical coherence times of up to
milliseconds \cite{Sun2002} and spin coherence times $>$ seconds
\cite{Longdell2005} have been demonstrated at low temperature
($\lesssim$ 4 K).

A quantum memory also requires a scheme for achieving efficient
and reversible mapping of the photonic qubit onto the atomic
ensemble. Techniques investigated include stopped light based on
electromagnetically induced transparency (EIT)
\cite{Chaneliere2005,Eisaman2005,Choi2008}, Raman interactions
\cite{Julsgaard2004,Nunn2007,Hosseini2009} or photon-echo based
schemes
\cite{Moiseev2001,Kraus2006,Hetet2008,Afzelius2009a,Riedmatten2008}.
Much progress has been made in terms of quantum memory efficiency
\cite{Choi2008,Simon2007a}, and storage time
\cite{Zhao2009,Zhao2009a}. Storage of multiple qubits is
challenging, however, because it requires a quantum memory that
can store many optical modes into which qubits can be encoded.
Note that each pair of modes can encode a different qubit, or more
generally, d modes encode a qudit. A mode can be defined in time \cite{Simon2007}, space \cite{Lan2009}, or frequency.  Time multiplexing as used in classical communication has the great advantage of requiring only a single spatial mode
\cite{Afzelius2009a,Nunn2008}, hence a single quantum memory.
Moreover, temporal modes can be used to define time-bin qubits
\cite{Gisin2007a}, which are widely used in fiber-based quantum
communication due to their resilience against polarization
decoherence in fibers.

This type of temporal multimode storage is difficult, however, due
to the scaling of the number of stored modes $N_m$ as a function
of optical depth $d$ of the storage medium
\cite{Afzelius2009a,Nunn2008}. For EIT and Raman interactions
$N_m$ scales as $\sqrt{d}$ \cite{Nunn2008}, making it very
difficult to store many modes. Recently we proposed
\cite{Afzelius2009a} a multimode storage scheme based on atomic
frequency combs (AFC) with high intrinsic temporal multimode
capacity \cite{Afzelius2009a,Nunn2008}. Using this method we
recently demonstrated \cite{Riedmatten2008} that a weak coherent
state $|\alpha\rangle_L$ with mean photon number
$\overline{n}=|\alpha|^2$ $<$ 1, can be coherently and reversibly
mapped onto a YVO$_4$ crystal doped with neodymium ions. Later
experiments \cite{chaneliere-2009,Afzelius2009b,Amari2009a} in
other RE-doped materials have improved the overall storage
efficiency (35\%) and storage time (20$\mu$s). Yet, in these
experiments at most 4 modes have actually been stored at the single photon level, thus the
predicted \cite{Afzelius2009a,Nunn2008} high multimode capacity
has yet to be shown experimentally.

\begin{figure*}[hbt]
\centering
\includegraphics[width=.95\textwidth]{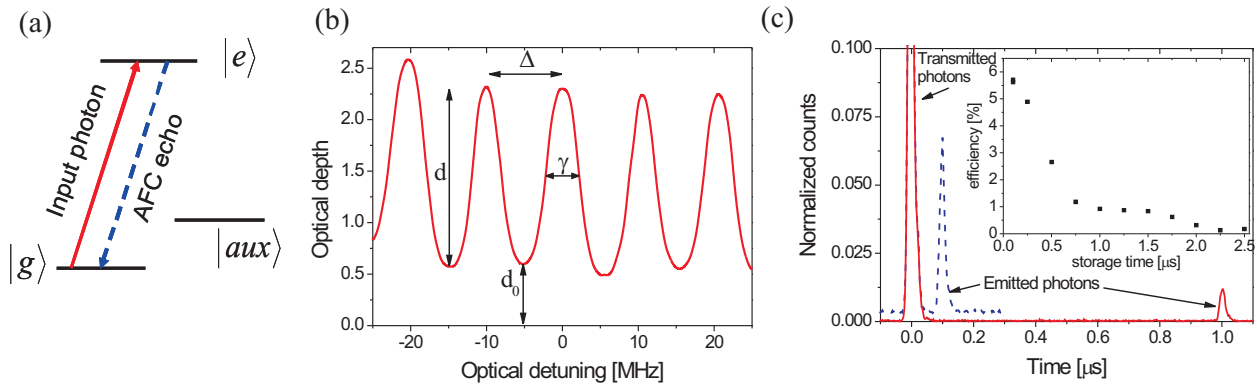}
\caption{(a) Simplified level scheme of the Nd ions doped into
Y$_2$SiO$_5$. We use the optical transition at 883nm between the
$^4$I$_\frac{9}{2}$ ground state and $^4$F$_\frac{3}{2}$ excited
state. The former is split into two Zeeman levels by a 0.3 Tesla
magnetic field ($|g\rangle$ and $|aux\rangle$). The experiment is
performed on $|g\rangle$-$|e\rangle$ where the absorption profile
is shaped into an AFC by optically pumping atoms into
$|aux\rangle$. The basic idea is to send in a pulse sequence on
$|g\rangle$ - $|e\rangle$ that has a periodic spectral density
(due its Fourier spectrum). Some of the excited atoms have a
certain probability to spontaneously de-excite to $|aux\rangle$.
The atoms left behind in $|g\rangle$ form the grating (see panel
(b)). To build up a deep grating the sequence is repeated many
times (up to the timescale of the population relaxation between
$|g\rangle$ and $|aux\rangle$). More details on the preparation
can be found in the Supplementary Information.
(b) An example of a generated comb with periodicity $\Delta$=10
MHz. The relevant AFC parameters defined in the text are
indicated. (c) Mapping of weak coherent states with
$\overline{n}=0.5$ (in a single temporal mode) onto the Nd-doped
crystal. Shown are two different experiments with $\Delta$=10 MHz
(dashed line offset vertically) and 1 MHz (solid line). The
photons that are transmitted without being absorbed are detected
at $t$=0, while the absorbed and re-emitted photons are detected
around $t=1/\Delta$. The vertical scale has been normalized such
that it yields efficiency. Inset: The overall write and read
efficiency as a function of $1/\Delta$.} \label{fig:AFC}
\end{figure*}

\section{Results}

Here we demonstrate reversible mapping of 64 temporal modes
containing weak coherent states at the single photon level onto
one atomic ensemble in a single spatial mode using an AFC-based
light-matter interface \cite{Afzelius2009a}. An AFC is based on a
periodic modulation (with periodicity $\Delta$) of the absorption
profile of an inhomogeneously broadened optical transition
$|g\rangle \rightarrow |e\rangle$ (see Fig. \ref{fig:AFC}). The
modulation should ideally consist of sharp teeth (with full-width
at half-maximum $\gamma$) having high peak absorption depth $d$,
cf. Figure \ref{fig:AFC}b. Such a modulation can be created by
optical pumping techniques (see Figure \ref{fig:AFC} and Methods).
This requires, however, an atomic ensemble with a static
inhomogeneous broadening and many independently addressable
spectral channels. This can be found in RE-doped solids where
inhomogeneous broadening is of order 1-10 GHz and the homogeneous
linewidth is of order 1-100 kHz when cooled $<$ 4K. When a weak
photonic coherent state $|\alpha\rangle_L$ with $\overline{n}<1$
is absorbed by the atoms in the comb, the state of the atoms can
be written as
$|\alpha\rangle_A=|G\rangle+\alpha|W\rangle+O(\alpha^2)$. Here
$|G\rangle=|g_1\cdot\cdot\cdot g_N \rangle$ represents the ground
atomic state and

\begin{equation}
|W\rangle=\sum_n c_n e^{i2\pi\delta_nt}e^{-ikz_n} |g\cdot\cdot\cdot e_n\cdot\cdot\cdot g \rangle
\label{eq_etatW}
\end{equation}

\noindent represents one induced optical excitation delocalized
over all the $N$ atoms in the comb. In Eq. (\ref{eq_etatW}) $z_n$
is the position of atom $n$, $k$ is the wave-number of the
single-mode light field, $\delta_n$ the detuning of the atom with
respect to the laser frequency and the amplitudes $c_n$ depend on
the frequency and on the spatial position of the particular atom
$n$. The initial (at $t$=0) collective strong coupling between the
light mode and atoms is rapidly lost due to inhomogeneous
dephasing caused by the $\exp(i2\pi\delta_nt)$ phase factors. If
we assume that the peaks are narrow as compared to the periodicity
(i.e. a high comb finesse $F=\Delta/\gamma$), then
$\delta_n\approx m_n\Delta$ and the W state will rephase after a
pre-programmed time $1/\Delta$. The rephased collective state W
will cause a strong emission in the forward direction (as defined
by the absorbed light).

\begin{figure} [hbt]
\centering
\includegraphics[width=.45\textwidth]{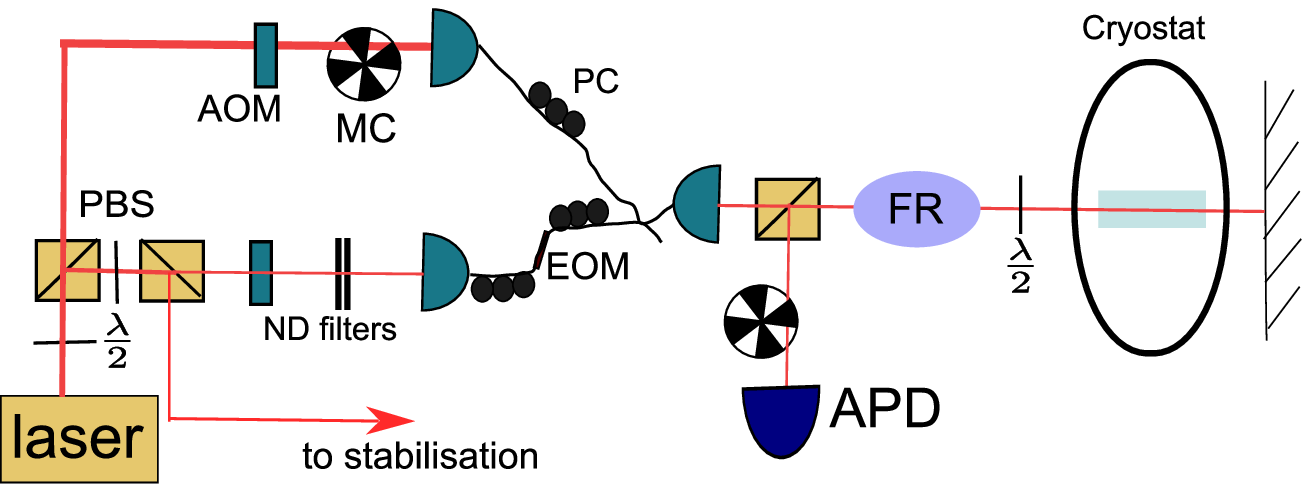}
\caption{The output from a frequency-stabilized ($<$100kHz) diode laser was split into two beams using a polarization beam splitter (PBS). Each beam could be amplitude, frequency and phase modulated using a double-pass acousto-optic modulators (AOM). One beam was used for creating the preparation pulses (see text), and
the other one for creating the weak pulses to be stored (strongly attenuated using neutral density (ND) filters). In the weak path an additional electro-optic amplitude modulator (EOM) was used to create short input pulses for the multimode storage
experiments. The paths were mode overlapped using a fiber-coupled beam combiner. The light was sent through the crystal, again in free space, in a double-pass setup using a Faraday rotator (FR) and a PBS. The output light was
collected with a multimode fiber and detected by an Si single-photon counter (APD). Two synchronized mechanical choppers (MC) blocked the detector during the preparation sequence and blocked the preparation beam during the storage sequence, respectively. See Supplementary Information for more details.} \label{fig:setup}
\end{figure}

This type of photon-echo emission is also
observed in accumulated or spectrally programmed photon echoes
\cite{Hesselink1979,Carlson1984,Mitsunaga1991,Schwoerer1994}, which inspired our proposal. Spectral atomic gratings have also been proposed \cite{Merkel1996} and demonstrated \cite{Tian2001} for coherent optical delay of streams of strong classical pulses. The interest in spectral gratings was recently renewed in the context of quantum memories, when it was realized how to achieve a much more efficient spectral grating than previously possible. This is possible due to the highly absorbing and sharp peaks in the AFC structure \cite{Afzelius2009a}. In practice the finite finesse of the
comb still needs to be accounted for, which causes a partial loss
of the collective state. But in Ref. \cite{Afzelius2009a} we
show theoretically that $F$=10 induces a negligible loss, which in
combination with a high optical depth $d$ makes the AFC scheme
very efficient. High-efficiency mapping using high-finesse combs
have been shown experimentally \cite{chaneliere-2009,Amari2009a}.
These experiments and the present work stores light for a
pre-determined time given by $1/\Delta$. We thus emphasize that we
also proposed \cite{Afzelius2009a}  and experimentally
demonstrated \cite{Afzelius2009b} a way to achieve on-demand
readout by combining AFC with spin-wave storage. On-demand readout is a crucial resource for applications in quantum networks in order to render different quantum channels independent.

The multimode property of an AFC memory can easily be understood
qualitatively. For a periodicity $\Delta$ and $N_p$ peaks, its
total bandwidth is of order $\sim N_p \Delta$ meaning that a pulse
of duration $\tau \sim 1/(N_p \Delta)$ can be stored. The
multimode capacity stems from the fact that the grating can absorb
a train of weak pulses before the first pulse is re-emitted after
$T=1/\Delta$ (cf. Fig. \ref{fig:AFC}c). This simple calculation
gives a multimode capacity $N_m \propto T/\tau \propto N_p$. Thus
a comb with many peaks $N_p$ allows us to create a highly
multimode memory in the temporal domain. In this context RE-doped
solids are particularly interesting due to their high spectral
channel density.

\begin{figure}[hbt]
\centering
\includegraphics[width=.50\textwidth]{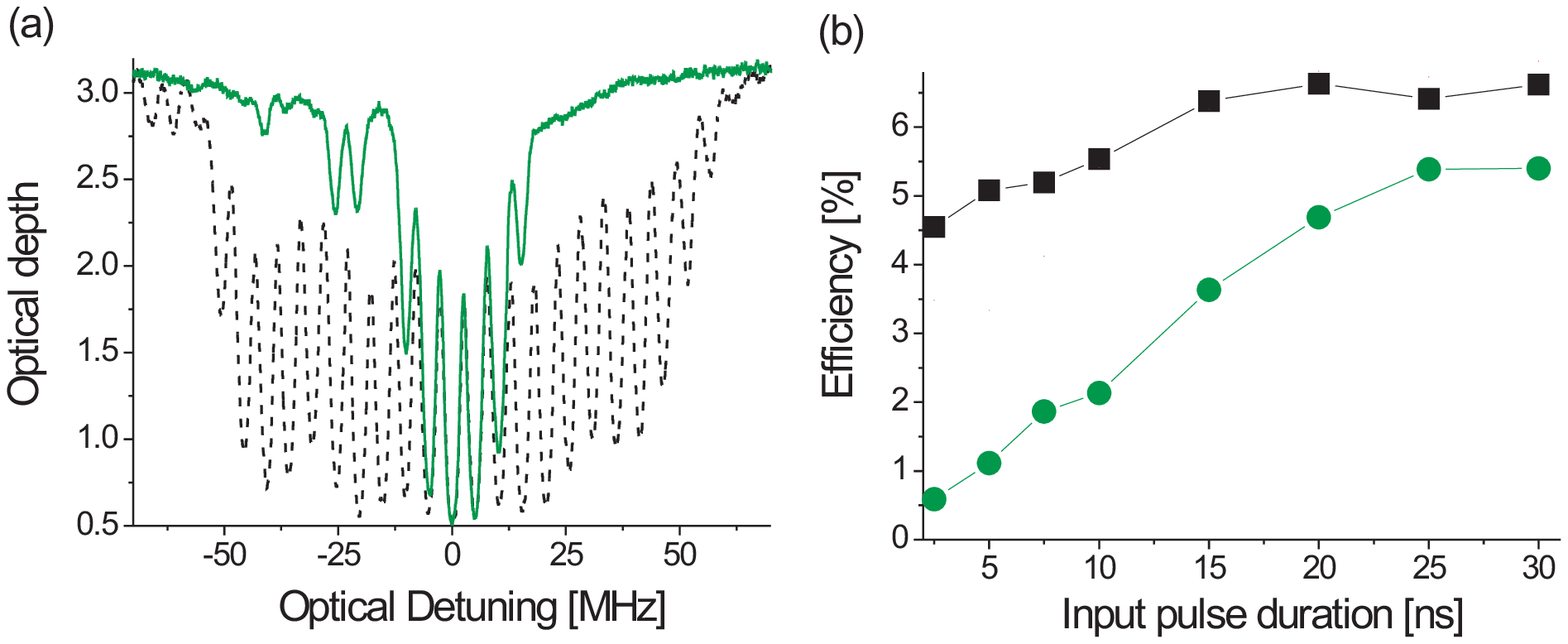}
\includegraphics[width=.40\textwidth]{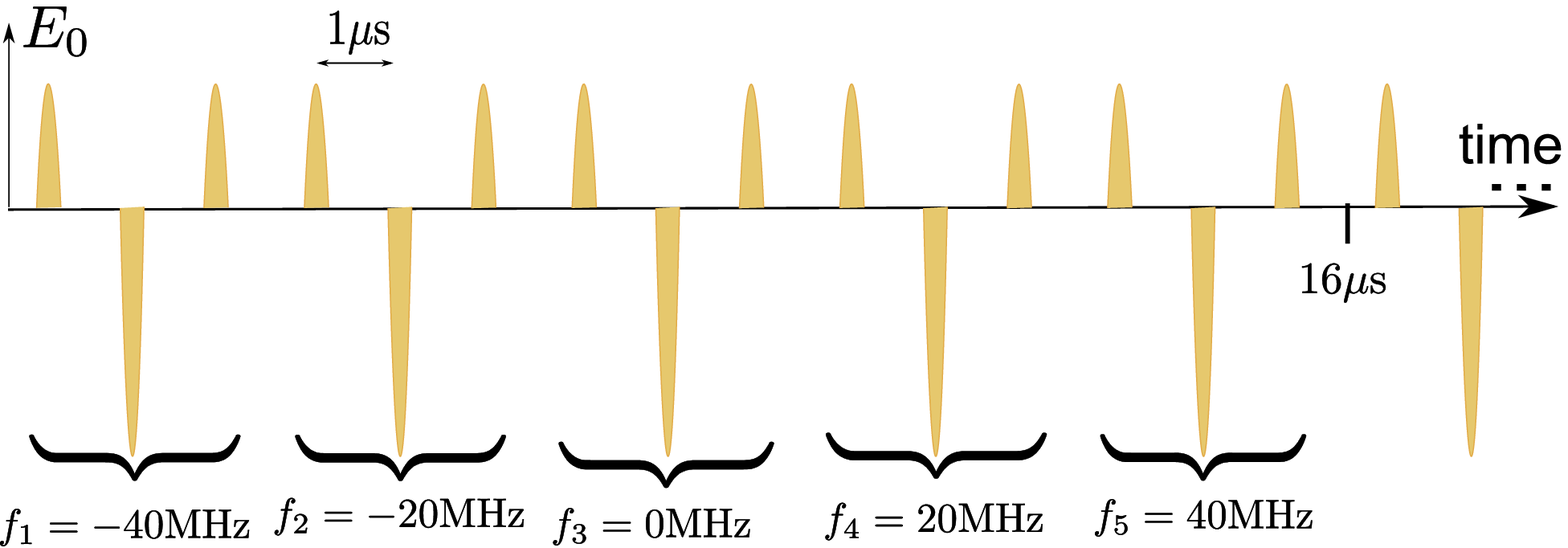}
\caption{(a) Experimental combs created using preparation
sequences with either single (solid line) or five (dashed line)
simultaneous pump frequencies. The frequency shifted sequences
allows us to enlarge the frequency range over which the optical
pumping is efficient and thereby creating a wide 100 MHz comb. (b)
Efficiency as a function of the duration (FWHM) of the input pulse
for a single (circles) and five (squares) frequency preparation.
As the duration decreases the bandwidth of the input pulse
increases. The decrease in efficiency for short pulses is due to
bandwidth mismatch for large bandwidths when using a single
preparation frequency. This experiment clearly illustrates the
gain in bandwidth in the extended preparation sequence for which
only a small decrease in efficiency is observed. (c) Pulse sequence for atomic frequency comb
preparation (see text). In order to increase the bandwidth, the pulses are
repeated with shifted frequencies. This pulse sequence was used
for most of our experiments. Here it creates a comb of 100MHz
bandwidth and a periodicity of 1MHz. The total sequence takes
16$\mu$s and it is repeated around 2000 times in order to prepare
the AFC.}
\label{fig:memory bandwidth}
\end{figure}

Here we work with a neodymium-doped Y$_2$SiO$_5$ crystal having a transition
wavelength at 883 nm with good coherence properties (see Methods
for the spectroscopic information). This wavelength is convenient
since we can work with a diode laser and Si based single photon
counters having low noise (300 Hz) and high efficiency (32\%).

\begin{figure*}[hbt]
\centering
\includegraphics[width=.90\textwidth]{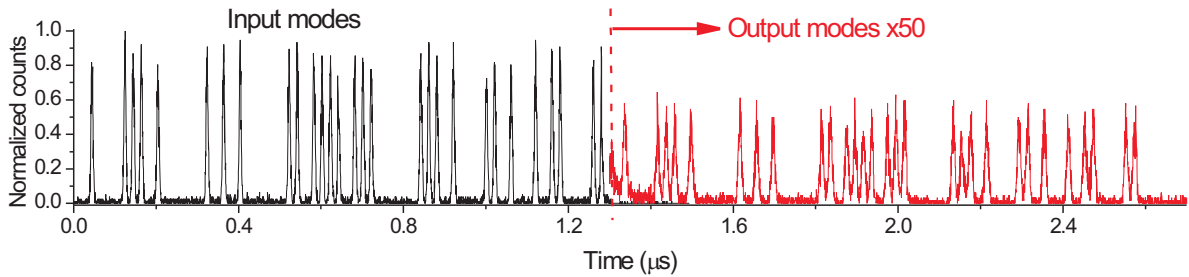}
\caption{Storage of an input state composed of 64 temporal modes.
The input (left part) is a random sequence of full and empty time
bins, where the mean photon number in the full ones is
$\overline{n}\lesssim$1. The output (right part) clearly preserves
the amplitude information to an excellent degree. The
pre-determined storage time was $T$=1.32 $\mu$s, the duration of
the input 1.28 $\mu$s, the mode separation 20 ns and the mode
duration (FWHM) 5 ns. The output has been multiplied with a factor
of 50 for clarity. The average storage and retrieval efficiency
was here 1.3\%. Other examples of multimode storage, e.g. with all
time bins filled can be found in the Supplementary Information.}
\label{fig:64modes}
\end{figure*}

The comb is prepared on the $|g\rangle$ - $|e\rangle$ transition by frequency selectively pumping atoms into an auxiliary state $|aux\rangle$ (see Fig. \ref{fig:AFC}). There are different techniques for achieving
this. For instance, by creating a large spectral hole, and then
transferring back atoms from an auxiliary state to create a comb,
as used in \cite{Afzelius2009b}. Here we use a similar technique
to \cite{Riedmatten2008}, where a series of pulses separated by a
time $\tau$ pump atoms from $|g\rangle$ to $|aux\rangle$ (through
$|e\rangle$) with a power spectrum having a periodicity
$1/\tau=\Delta$. This technique is also frequently used in accumulated photon echo techniques \cite{Hesselink1979,Tian2001}. Here each pulse sequence consisted of three pulses where the central pulse is $\pi$-dephased, which has a
power spectrum with "holes" (see Fig. \ref{fig:memory bandwidth}c). A straightforward calculation shows
that the width of the holes in the power spectrum decreases with
the number of pulses in the sequence. In this experiment three
pulses were enough to reach the optimal comb finesse (F$\approx$3) to achieve the maximal efficiency for a given optical depth (see Methods). Note that
the sum of the amplitudes of the side pulses (here two) should
correspond to the amplitude of the central $\pi$-dephased pulse,
in order to obtain the appropriate power spectrum. This rule also
holds for sequences with more pulses. To increase the depth of the
comb the sequence was repeated 2000 times. More details can be found in the
Supplementary Information.

The experimental sequence is divided into two parts: the
preparation of the AFC (cf. above) and the storage of the weak pulses. The
preparation lasts 100 ms, which is followed by a delay of 5 ms( $\approx 17
T_1$) to avoid fluorescence noise from atoms left in the excited state. During the storage sequence, 1000 independent
trials are performed at a repetition rate of 200 kHz. The entire
sequence preparation plus storage is then repeated with a
repetition rate of 5 Hz. An overview of the experimental set up is shown in Fig. \ref{fig:setup}.

In Fig. \ref{fig:AFC}c we show storage experiments with pre-determined storage times of
$T$=100ns and 1$\mu$s, for a single temporal mode. The overall
in-out mapping efficiencies, defined as the ratio of the output
pulse counts to the input pulse counts, are $\sim 6\%$ and $\sim
1\%$ respectively (see inset of Fig. \ref{fig:AFC}c). In the
Methods section we present a theoretical analysis of the
efficiency performance. The efficiency for single-mode storage is
currently lower than has been achieved in the best-performance
single-mode memories, e.g.
\cite{Choi2008,Hetet2008,Hosseini2009,chaneliere-2009,Amari2009a}.
But as explained later our interface compares very favorably to
these experiments in terms of potential multimode storage
efficiency.

The main goal of the present work is to demonstrate high multimode
storage, as theoretically predicted in
\cite{Afzelius2009a,Nunn2008}. Following the discussion above, we
should maximize the number of peaks in the comb. This can be done
by increasing the density of peaks in a given spectral region
(i.e. increasing the storage time $T$) or by changing the width of
the AFC (i.e. increasing the bandwidth). Here we fix the storage
time to $T$=1.3 $\mu$s where we reach an efficiency of $\gtrsim$
1\% and concentrate our efforts on increasing the bandwidth. The
spectral width of the grating is essentially given by the width of
the power spectrum of the preparation sequence, which here results
in a width of about 20-30 MHz. We can however substantially
increase the total width by inserting more pulses in the
preparation sequence, which are shifted in frequency (see
Methods). We thus optically pump atoms over a much larger
frequency range. Note that the frequency shift should be a
multiple of $\Delta$ in order to form a grating without
discontinuities. In this way we managed to extend the bandwidth of
the interface to 100 MHz as shown in Fig. \ref{fig:memory
bandwidth}a, without significantly affecting the AFC echo
efficiency. This is illustrated in Figure \ref{fig:memory
bandwidth}b, where we show storage efficiency as a function of the
duration of the input pulse when the preparation sequence contains
a single or five frequencies. The maximum bandwidth allows us to
map short $<$10 ns pulses into the memory.

\begin{figure}[b]
\centering
\includegraphics[width=.50\textwidth]{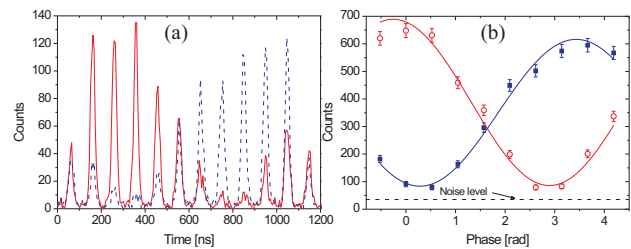}
\caption{The coherence of the multimode storage was measured via
an interference measurement. (a) The output signal (solid line)
generated by the double-AFC scheme (see text), which causes an
interference between consecutive modes. The input sequence (not
shown) is a series of weak coherent states
($\overline{n}\approx$1.8) of equal amplitude ($c_k^2=c_m^2$)
where the relative phase difference between modes
$\phi_{k+1}-\phi_k$ ranges from -$\pi$/6 to 8$\pi$/6 with a step
of $\pi$/6. This allows us to capture a complete interference
fringe in one measurement. It also clearly demonstrates the
preservation of coherence over the complete multimode output. By
changing the detuning of the centre of the second AFC with respect
to the carrier frequency of the light, we can impose an additional
relative $\pi$ phase \cite{Afzelius2009a} on the corresponding
output (see Supplementary Information). This shifts the
interference fringe with half a period (dashed line). (b) The
corresponding net interference visibilities are 86\%$\pm$3 (open
circles) and 85\%$\pm$3\% (filled squares), with detector noise
(dashed line) subtracted. The uncorrected raw visibilities were
78\%$\pm$3\% and 76\%$\pm$2\%.} \label{fig:interference}
\end{figure}

In addition to the present motivation for multimode storage, a
large bandwidth is equally interesting for interfacing a memory
with non-classical single-photon or photon pair sources. These
usually have large intrinsic bandwidth which requires extensive
filtering for matching bandwidths. In the present case our
extended bandwidth ($\times$5) would require a corresponding
factor of less filtering.

We show the high multimode capacity of our interface by storing 64
temporal modes during a pre-determined time of 1.3 microseconds
(see Fig. \ref{fig:64modes}), with an overall efficiency of 1.3\%.
This capacity is more than an order of magnitude higher than
previously achieved for multiplexing a quantum memory in a single
spatial mode \cite{Riedmatten2008,Hosseini2009}. As shown we can
store a random sequence of weak coherent states. Storage of random
trains of single photon states has been proposed for multiplexing
long-distance quantum communication systems based on so called
quantum repeaters \cite{Duan2001,Sangouard2009a}. The maximum rate
of communication would then be proportional to the number of modes
that can be stored \cite{Simon2007}. Our experiment clearly shows
the gain that can be made using an AFC-based quantum memory. It
thus opens up a route towards achieving efficient quantum
communication using quantum repeaters.

It is now possible to use consecutive temporal modes, e.g. modes
$|k\rangle$ and $|k+1\rangle$, to encode time-bin qubits
$c_k|k\rangle + c_m e^{i\phi_{km}}|m\rangle$, in which case a good
coherence between modes is crucial. The coherence can be measured
by preparing superposition states and performing projective
measurements using an interferometric set up. Projective
measurements on time-bin qubits is usually performed using an
unbalanced Mach-Zehnder interferometer (MZI) where consecutive
time-bin are interfering \cite{Gisin2007a}. We can perform the
same task with our light-matter interface by using a double-AFC
scheme (with $\Delta_1$ and $\Delta_2$) as shown in
\cite{Riedmatten2008}. In short, the difference in delay
$1/\Delta_1-1/\Delta_2$ plays the role of the delay in a
unbalanced MZI. We observe excellent coherence over all modes with
an average visibility of $V=86\%\pm3\%$, see Fig.
\ref{fig:interference}, corresponding to a conditional qubit
fidelity of $F=(1+V)/2\approx93\%$.

\begin{figure}[hbt]
\centering
\includegraphics[width=.45\textwidth]{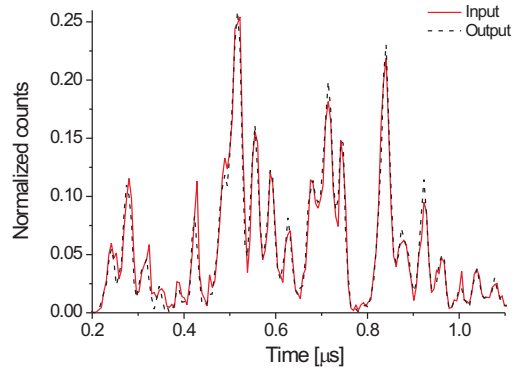}
\caption{Mapping of a 1 $\mu$s long input pulse with randomly
varying amplitude. As seen the overlap between the normalized
input (dashed line) and output (solid line) pulses is excellent.
The total average number of photons in the input pulse is
$\overline{n}\approx4$}. \label{fig:arbitrary}
\end{figure}

To further illustrate our ability to store multimode light states
we create a light pulse with a random amplitude modulation. As
shown in Fig. \ref{fig:arbitrary} we can faithfully store this
kind of light pulses. The possibility of storing weak arbitrary
light states using photon-echo based schemes was pointed out
already by Kraus et al. \cite{Kraus2006}. We believe that this
work, where complex phase and amplitude information are reversible
and coherently mapped onto one atomic ensemble, is the first
experimental realization showing these properties at the single
photon level.

\section{Discussion}

For multimode storage the efficiency of our
interface would outperform the current EIT and Raman based quantum
memories in homogeneously broadened media, although impressive
efficiencies have been achieved for single-mode storage
\cite{Choi2008,Simon2007a,Hammerer2008}. This is due to the poor
scaling of the efficiency as a function of the number of modes for
a given optical depth \cite{Nunn2008}. It also compares favorably
to the recent few modes storage experiment \cite{Hosseini2009}
using the gradient echo memory (GEM), another echo based storage
scheme, also due to the scaling of mode capacity for a given
optical depth (N$_m\sim d$)\cite{Simon2007,Nunn2008}. Still, an
increase in storage efficiency and on-demand read out is necessary
for applications in quantum communication.

The next grand challenge is to combine multimode storage, high efficiency
\cite{Amari2009a} and on-demand read out \cite{Afzelius2009b} in
one experiment. The immediate efforts will most probably by devoted to
praseodymium and europium-doped Y$_2$SiO$_5$ crystals, where the ground state manifold has the necessary number of spin levels (three levels) for implementing the on-demand readout. The recent achievements in Pr-doped Y$_2$SiO$_5$ crystals are very encouraging \cite{Afzelius2009b,Amari2009a}, although the bandwidth was limited to a few MHz due to the hyperfin level splitting. Europium-doped Y$_2$SiO$_5$ has the potential of offering higher bandwidths (up to 70 MHz) and narrower comb peaks, which results in higher multimode capacity \cite{Afzelius2009a}. In order to exploit the high-bandwidth results reported in this work, using neodymium-doped crystals, one needs to find a third spin level with a long spin coherence lifetime. An interesting path forward is to investigate neodymium isotopes with a hyperfine structure ($^{143}$Nd and $^{145}$Nd) \cite{Macfarlane1998}. Recent results on a similar system \cite{Bertaina2007}, $^{167}$Er$^{3+}$:CaWO$_4$, show coherence times approaching 100 $\mu$s for hyperfine transitions. Clearly this path requires extensive spectroscopic studies in order to optimize the spin population and coherence lifetimes. But it is very interesting since it opens up several material candidates (e.g. doped with Erbium \cite{Lauritzen2009} and Neodymium) for quantum memory applications.

To summarize we have demonstrated the reversible mapping of up to
64 optical temporal modes at the single photon level onto one
solid state atomic ensemble. We have demonstrated that the quantum
coherence of the stored modes is preserved to a high extent. The
different modes can then be used to encode multiple time-bin
photonic qubits. Alternatively, they could also be considered as
high-dimensional qudits states. This opens up possibilities to
store higher dimensional quantum states such as entangled qudits
encoded in time bin bases. Our experiment opens the way to
multi-qubit quantum memories, which are a crucial requirement for
realistic quantum networks.

\section{Methods}

\footnotesize{
\textbf{Sample}

The sample is a 10mm long neodymium-doped yttrium orthosilicate
crystal(Nd$^{3+}$:Y$_2$SiO$_5$) with a low Nd$^{3+}$ concentration
of 30 ppm. The inhomogeneous broadening of the
$^4$I$_\frac{9}{2}$-$^4$F$_\frac{3}{2}$ absorption line is around
6GHz and the optical depth 1.5 for this sample. By using a
double-pass set up through the crystal we could increase the
optical depth to 3. We measured an excited state lifetime of $T_1$=300$\mu$s using fluorescence
spectroscopy and stimulated photon echoes. With conventional
photon echoes (two-pulse) we measure a homogeneous linewidth of
3.4kHz (T$_2=92.7\mu$s). Each level is a Kramer's doublet which
split into two spin states in a magnetic field. For the field
orientation used in this experiment we measured g factors of
$g_g=2.6$ and $g_e=0.5$. In a 300mT magnetic field, the excited
states were separated with 2GHz. We measured a ground state Zeeman
population relaxation lifetime, by spectral hole burning (SHB), of
around $T_{1Z}$=100ms. In the SHB measurements we also observed a
superhyperfine interaction of Nd ions with yttrium. This causes
additional spectral side holes at around 640kHz (for the present
magnetic field), thus the effective homogeneous linewidth is
around 1 MHz. This was our main limitation for the efficiency of
our light-matter interface since it affected our ability to create
a good comb for the longer storage times ($1/\Delta\approx1\mu$s).

\textbf{Storage efficiency analysis}

The efficiency can be calculated theoretically using the formula
\cite{Afzelius2009a,Riedmatten2008} $\eta \approx (d/F)^2 e^{-d/F}
e^{-7/F^2} e^{-d_0}$. The different terms can be given a
qualitative understanding. The first term represents the
collective coupling, the second the re-absorption of the
re-emitted light, the third is an intrinsic dephasing factor due
to the finesse and the last term a loss due to an absorption
background $d_0$. For the comb with $\Delta$=10MHz we measure
$d\approx1.7$,$F\approx2.7$ and $d_0\approx0.5$ (see Fig.
\ref{fig:AFC}b), resulting in a theoretical efficiency of $\eta
\approx5\%$ in close agreement with the experiment (see Fig.
\ref{fig:AFC}c). The major limiting factor here is $d_0$ (caused
by imperfect preparation of the comb) and then the optical depth
of the comb $d$ (the finesse being close to optimum for this $d$
\cite{Afzelius2009a}). The decrease in efficiency for longer
storage times (see inset of Fig. \ref{fig:AFC}c) is principally
due to an increase in the background absorption $d_0$ and an
accompanying decrease in the peak absorption $d$. This in turn is
caused by the effective spectral resolution of 1 MHz in the
optical pumping, which is a limitation of the present material
(see Methods above). However, the storage efficiency is between
one and two orders of magnitude higher than what we achieved in
the material Nd:YVO$_4$ \cite{Riedmatten2008}, which we attribute
to an improvement in optical pumping in this Nd-doped material.

\textbf{Acknowledgements} We acknowledge financial support from
the Swiss NCCR Quantum Photonics, the EC projects Qubit
Applications (QAP) and ERC Advanced Grant (QORE). We also
acknowledge useful discussions with Christoph Simon and Nicolas
Sangouard.

\newpage

\appendix
\section{Supplementary Information}

We provide additional information for our manuscript. This
material includes a detailed description of the experimental setup
and of the atomic frequency comb preparation, as well as
additional results.

\section{Experimental details}

\subsection{Setup}

The experimental setup is shown in Fig.\ref{fig:setup}. The light
is emitted by an external cavity diode laser (Toptica) at 883nm
and is then split with a polarizing beam splitter (PBS) into two
paths. In both paths, the amplitude, frequency and phase of the
light can be modulated using acousto-optic modulators (AOM,
AA-Opto-Electronics) in double pass configuration. The path on the
top is used to prepare an AFC into the sample. To achieve this,
the AOM creates a series of pulses with a characteristic spectrum
explained in the next section. The other path is used to create
the weak pulses of light at the single photon level that are going
to be mapped onto the crystal. With an AOM and neutral density(ND)
filters we create pulses containing less than one photon in
average. The AOMs have a good extinction ratio ($>$ 40 dB) and
allows us to shift the frequency or the phase of the light.
However, to create very short pulses(few ns), we  use an
additional a 20 GHz electro-optical modulator(EOM, EO Space).

\begin{figure} [hbt]
\centering
\includegraphics[width=.45\textwidth]{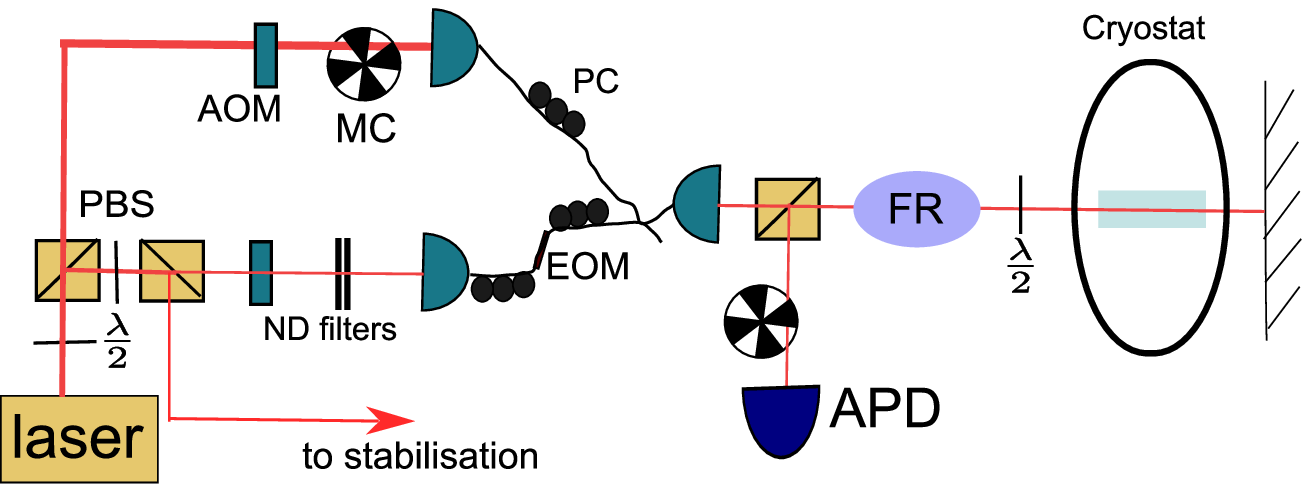}
\caption{\footnotesize{
Experimental setup : see text for details. The following abbreviations are used : polarizing beam splitter (PBS),  acousto-optic modulator (AOM), neutral density filter(ND filter), mechanical chopper (MC), polarization controller (PC), electro-optical modulator (EOM), Faraday rotator (FR) and avalanche photo-diode (APD).}} \label{fig:setup}
\end{figure}

Both path are then combined using a 90/10 fiber coupler and sent
into free space into the sample cooled at 3K using a pulse tube
refrigerator (Oxford Instruments). The PBS combined to a $\lambda
/2$ wave-plate allow us to adjust the polarization of the light
parallel to the D1 crystallographic axis (optical axis). In order
to increase the available optical depth, we implemented a
double-pass setup using a Faraday Rotator(FR). In this way, we
have a configuration where the polarization of the light remains
constant when propagating in double pass through the crystal, but
is rotated by 90 degrees after the second pass in the FR, such
that the the light emitted by the crystal can be separated from
the input light, thanks to a PBS.

The light is then coupled to a multimode optical fiber and
detected by a single photon Si Avalanche Photo-Diode (APD) with
32\% efficiency. The mechanical chopper(MC) in the preparation
path is used to block the leakage of the preparation AOM when the
APD is detecting echoes. The second MC is used to protect the APD
when strong preparation pulses are used. The transmission between
the input of the cryostat and the APD was typically between 25$\%$
and 30 $\%$. Finally, a third path (partially shown), is used to
actively frequency stabilize the laser to less than 100kHz using
the Pound-Drever-Hall technique. The frequency reference is given
by a spectral hole in the crystal. The stabilization light is sent
into the same sample, but with a slightly different angle.

The experiment is repeated every 200ms, allowing sufficient time
for the ground state population to relax between experiments. The
first 100ms were used for the preparation sequence to create an
AFC. Then, we waited 5ms to avoid fluorescence during the storage
and retrieval sequence, due to atoms left in the excited state
after the preparation of the AFC. After that, we perform $\sim$
1000 independent storage trials separated by 5$\mu$s. Each trial
contains the particular pulse sequence to be stored and the
re-emitted AFC echoes. Depending on the mean photon number per
pulses, the total integration time to accumulate sufficient
statistics varied from 30s to 10min.

\subsection{Comb Preparation}

We now explain in more detail the preparation sequence allowing us
to create a desired AFC. Similarly to \cite{deRiedmatten2008}, we
send series of pulses separated by a time $\tau$, which create a
periodicity of $\Delta=1/\tau$ in the absorption profile. By
sending repeatedly such sequence, this modulation will take the
form of a comb with sharp peaks. However, instead of sending only
pairs of pulses as in \cite{deRiedmatten2008}, we extended the
method by allowing the possibility of sending N pulses. The
spectrum of this series of N pulses is a a comb of periodicity
$\Delta=1/\tau$. However, we want to remove from the initial state
only atoms that are not in the desired comb. The spectrum of the
sequence must thus be the inverse of a comb : a series of holes
separated by $\Delta$. To achieve this, the central pulse must be
$\pi$-dephased, and with a pulse area equal to the sum of the
other pulses (see Fig.2).

The property of the comb can be determined by the characteristic
of the pulses sequence. As already mentioned, the periodicity
$\Delta$ is given by the time $\tau$ between pulses. The duration
of one pulse(or its spectrum) will determine the bandwidth of the
whole comb. Finally, the width of each peak will be inversely
proportional to the duration of the whole pulse sequence.
Similarly, we can say that the finesse of the comb will increase
with the number of pulses (F$\sim$ N$_{\mathrm{pulses}}$). Note
that these rules are true for the spectrum of the light, but the
process of spectral hole burning being more complex, the property
of the AFC can be different. For example, the width of each peak
is limited by the crystal properties, such as homogeneous
linewidth and superhyperfine interaction.

The comb preparation sequence for most of our experiments is shown
in Fig.\ref{fig:preparation}. It takes 16$\mu$s and it is repeated around 2000 times. The available optical depth being
relatively small (d=3), the optimal finesse to get the maximum
storage end retrieval efficiency was about 3 \cite{Afzelius2009}.
It was thus sufficient to send only series of three pulses (N=3)to
create the desired comb. The spectrum of the comb created in this
way was around 20MHz. It was possible to extend this bandwidth by
sending other series of pulses at a shifted
frequency($\pm$20MHz,$\pm$40MHz,etc\dots). Since there is no
interference between different frequencies, the pulses can be sent
within the coherence time $T_2$. This means that the whole
preparation sequence take the same time as with a single
frequency, and does not require more Rabi frequency.

\begin{figure}[hbt]
\centering
\includegraphics[width=.5\textwidth]{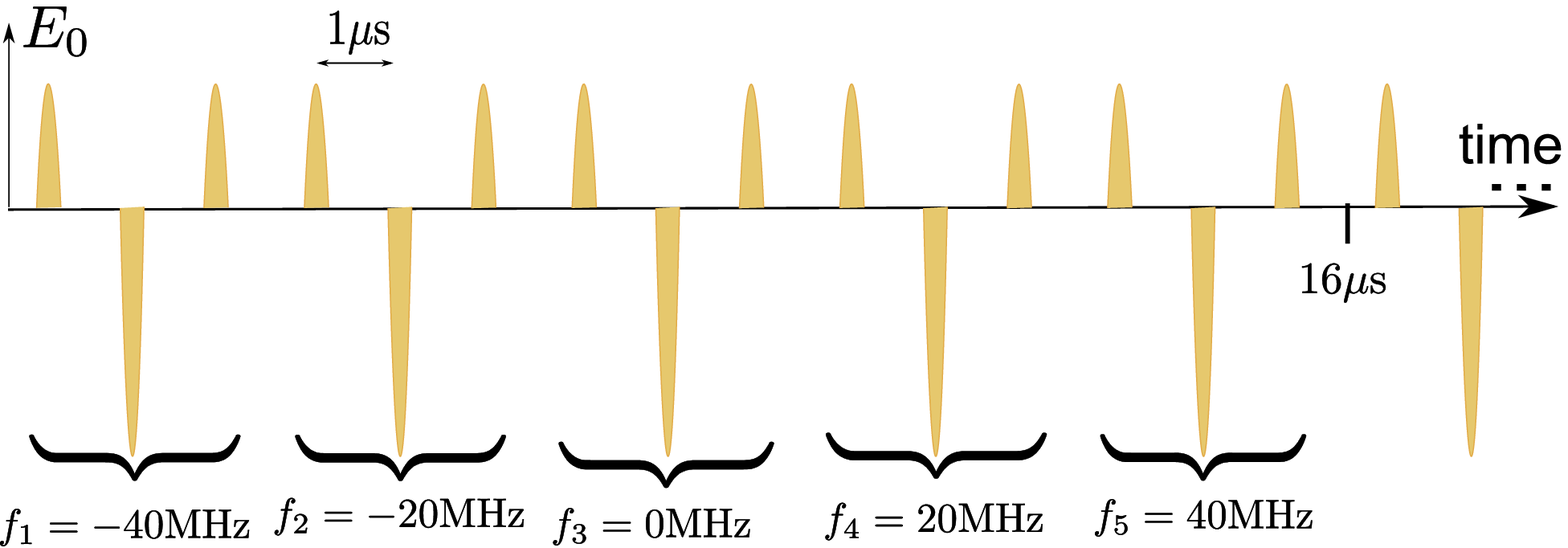}
\caption{\footnotesize{Pulse sequence for atomic frequency comb
preparation. In order to increase the bandwidth, the pulses are
repeated with shifted frequencies. This pulse sequence was used
for most of our experiments. Here it creates a comb of 100MHz
bandwidth and a periodicity of 1MHz. The total sequence takes
16$\mu$s and it is repeated around 2000 times in order to prepare
the AFC.}} \label{fig:preparation}
\end{figure}

\subsection{Double read-out using double AFC}
For the interference experiment, the stored excitations must be
read-out at different times. For that purpose, we need to create
multiple AFCs, with different periodicity $\Delta_i$ and possibly
different phases. This can be realized by using a preparation
pulse sequences with different pulse separations. Suppose that we
want to create two AFCs with periodicity $\Delta_1$ and
$\Delta_2$. The necessary sequence  is shown in
Fig.\ref{fig:preparation2AFC}. We send a pulse sequence with N
pulses separated by $\tau_1$ (with the central pulse at t=0),
superposed with N pulses separated by $\tau_2$ (with the central
pulse also at t=0). Thus, the total number of pulses will be 2N-1,
and the pulse area of the central one is equal to the sum of the
others. When we send a storage pulse in the sample, two echo will
be emitted, after a time $\tau_1$ and $\tau_2$. In our case, we
use N=3, such that the total number of pulses for creating two
AFCs with different periodicity is 5 (for one pumping frequency).
Note that when we superpose two combs, some peaks of the two AFCs
can be at the same position and be summed. This means that some
peaks will be higher than others. Thus, using this method  the two
AFCs cannot be created independently (i.e. with two independent
pulses series) or the absorption profile will not correspond to
the sum of two AFCs and we will face additional echoes. So we must
create directly two AFCs in one sequence.

Finally, we would like to be able to induce a phase $\phi$ in one
of the AFCs, in other world to shift the frequency of the comb
\cite {Afzelius2009}. To do so, we must add a phase in each pulse
preparing the corresponding comb (See Fig. 3). If we label the
pulses on the right of the central one with k=1,2,3\dots and the
pulses on the left with k=-1,-2,-3\dots, then the phase in the
pulse k must be k$\phi$.

\begin{figure}[hbt]
\centering
\includegraphics[width=.5\textwidth]{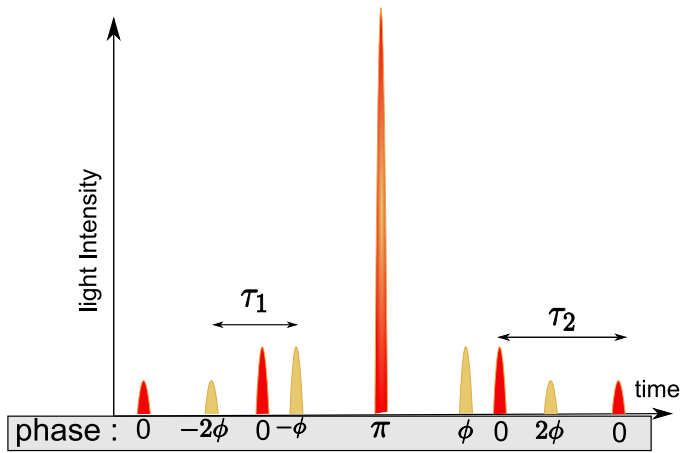}
\caption{\footnotesize{Pulse sequence for the preparation of two
atomic frequency combs, as required for the double read out for
the interference experiments. In this example, the two AFCs have a
periodicity $1/\tau_1$ and $1/\tau_2$. The first AFC will also get
a phase $\phi$.}} \label{fig:preparation2AFC}
\end{figure}

\section{Additional results}

We present here some additional results that illustrate the
multimode capacity of AFC. In Fig. \ref{fig:combs} we show
experimental combs with 1MHz peak separation (corresponding to 1
$\mu s$ storage time) and 20 MHz and 100 MHz bandwidth,
respectively. As noted, before the enlargement of the bandwidth
does not affect the center of the comb. Thus it allows us to
increase the number of mode for a fixed storage time without
affecting the efficiency (see Fig.2 of the manuscript). However,
compared to combs with bigger peak separation(shorter storage
time), the height of the peaks has decreased, and the absorption
background has increased, which explained the decay of efficiency
with storage time. As explained in the paper, the number of mode
is proportional to the number of peaks in the comb. Here we create
more than 100 peaks which illustrate the multimode performance of
our experiments.

\begin{figure}[hbt]
\centering
\includegraphics[width=.5\textwidth]{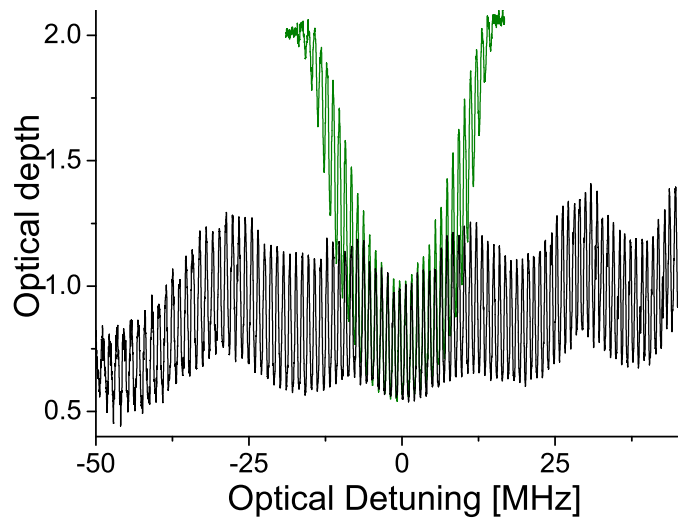}
\caption{\footnotesize{Experimental atomic frequency combs with 1
MHz peak separation for single(green line) or five(black line)
different simultaneous pump frequencies.}} \label{fig:combs}
\end{figure}

Finally, we show two more example of storage of 64 temporal modes
with different inputs. In Fig.\ref{fig:64modesbis}, all the modes
are full, which allows us in principle to store 32 time-bin
qubits. In Fig.\ref{fig:64modesbis2}, we modulated the amplitude
of the input pulses. We note that even if the input mode changes,
we do not need to adapt the AFC preparation. Indeed, the
efficiency is constant for all modes, and the echo always closely
follow the input pulses.

\begin{figure*}[hbt]
\centering
\includegraphics[width=\textwidth]{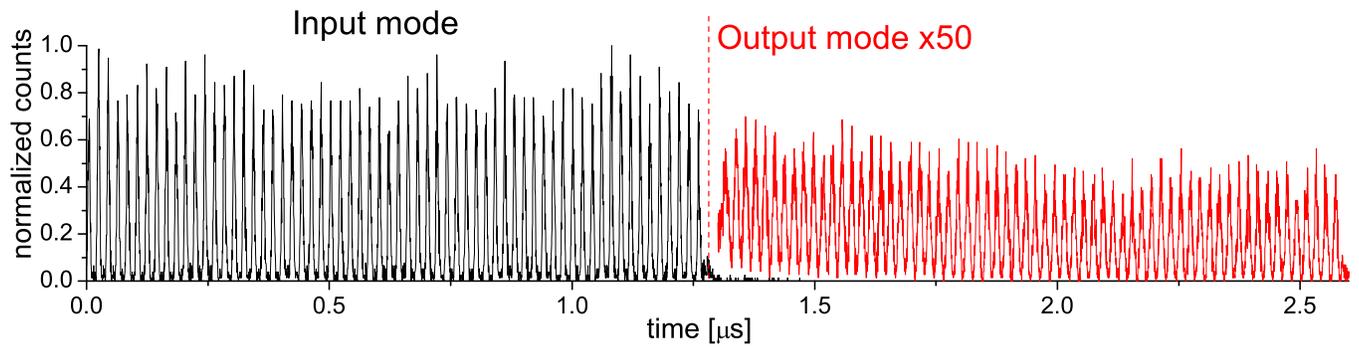}
\caption{\footnotesize{Storage of 64 consecutive pulses.
An efficiency of 1.4\% was measured. The mean photon number per pulse is $\bar{n}\approx 0.5$}}
\label{fig:64modesbis}
\end{figure*}
\begin{figure*}[hbt]
\centering
\includegraphics[width=\textwidth]{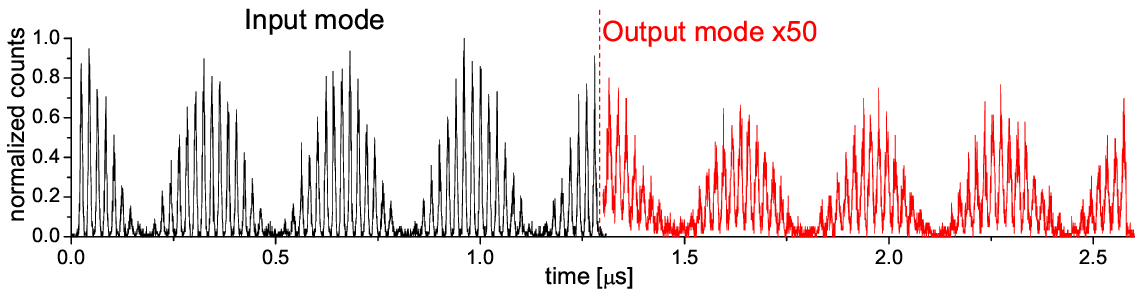}
\caption{\footnotesize{Storage of 64 pulses with a modulating amplitude.
An efficiency of 1.6\% was measured. The mean photon number in the biggest pulses is $\bar{n}\approx 0.9$}}
\label{fig:64modesbis2}
\end{figure*}

\bibliographystyle{unsrt}

\begin{thebibliography}{10}

\bibitem{Gisin2007a}
Gisin, N. and Thew, R.
\newblock Quantum communication.
\newblock {\em Nat Photon}{ \bf 1}, 165 (2007).

\bibitem{Gisin2002}
Gisin, N., Ribordy, G., Tittel, W., and Zbinden, H.
\newblock Quantum cryptography.
\newblock {\em Rev. Mod. Phys.}{ \bf 74}, 145 (2002).

\bibitem{Bennett1993}
Bennett, C.~H., Brassard, G., Cr\'{e}peau, C., Jozsa, R., Peres,
A., and Wootters,
  W.~K.
\newblock Teleporting an unknown quantum state via dual classical and
  einstein-podolsky-rosen channels.
\newblock {\em Phys. Rev. Lett.}{ \bf 70}, 1895 (1993).

\bibitem{Kimble2008}
Kimble, H.~J.
\newblock The quantum internet.
\newblock {\em Nature}{ \bf 453}(7198), 1023 (2008).

\bibitem{Duan2001}
Duan, L.-M., Lukin, M.~D., Cirac, J.~I., and Zoller, P.
\newblock Long-distance quantum communication with atomic ensembles and linear
  optics.
\newblock {\em Nature}{ \bf 414}, 413 (2001).

\bibitem{Sangouard2009a}
Sangouard, N., Simon, C., de~Riedmatten, H., and Gisin, N.
\newblock Quantum repeaters based on atomic ensembles and linear optics.
\newblock {\em arXiv:0906.2699} (2009).

\bibitem{Simon2007}
Simon, C., de~Riedmatten, H., Afzelius, M., Sangouard, N., Zbinden, H., and
  Gisin, N.
\newblock Quantum repeaters with photon pair sources and multimode memories.
\newblock {\em Phys. Rev. Lett.}{ \bf 98}, 190503 (2007).

\bibitem{Hammerer2008}
Hammerer, K., S{\o}rensen, A., and Polzik, E.
\newblock Quantum interface between light and atomic ensembles.
\newblock {\em arXiv:0807.3358} (2008).

\bibitem{Sun2002}
Sun, Y., Thiel, C.~W., Cone, R.~L., Equall, R.~W., and Hutcheson, R.~L.
\newblock Recent progress in developing new rare earth materials for hole
  burning and coherent transient applications.
\newblock {\em J. Lumin.}{ \bf 98}, 281 (2002).

\bibitem{Longdell2005}
Longdell, J.~J., Fraval, E., Sellars, M.~J., and Manson, N.~B.
\newblock Stopped light with storage times greater than one second using
  electromagnetically induced transparency in a solid.
\newblock {\em Phys. Rev. Lett.}{ \bf 95}, 063601 (2005).

\bibitem{Chaneliere2005}
Chaneli\`{e}re, T., Matsukevich, D.~N., Jenkins, S.~D., Lan, S.-Y., Kennedy, T.
  A.~B., and Kuzmich, A.
\newblock Storage and retrieval of single photons transmitted between remote
  quantum memories.
\newblock {\em Nature}{ \bf 438}, 833 (2005).

\bibitem{Eisaman2005}
Eisaman, M.~D., Andr\'{e}, A., Massou, F., Fleischhauer, M., Zibrov, A.~S., and
  Lukin, M.~D.
\newblock Electromagnetically induced transparency with tunable single-photon
  pulses.
\newblock {\em Nature}{ \bf 438}, 837 (2005).

\bibitem{Choi2008}
Choi, K.~S., Deng, H., Laurat, J., and Kimble, H.~J.
\newblock Mapping photonic entanglement into and out of a quantum memory.
\newblock {\em Nature}{ \bf 452}, 67 (2008).

\bibitem{Julsgaard2004}
Julsgaard, B., Sherson, J., Cirac, J.~I., Fiur\'{a}\v{s}ek, J., and Polzik,
  E.~S.
\newblock Experimental demonstration of quantum memory for light.
\newblock {\em Nature}{ \bf 432}, 482 (2004).

\bibitem{Nunn2007}
Nunn, J., Walmsley, I.~A., Raymer, M.~G., Surmacz, K., Waldermann, F.~C., Wang,
  Z., and Jaksch, D.
\newblock Mapping broadband single-photon wave packets into an atomic memory.
\newblock {\em Phys. Rev. A}{ \bf 75}, 011401 (2007).

\bibitem{Hosseini2009}
Hosseini, M., Sparkes, B.~M., H\'{e}tet, G., Longdell, J.~J., Lam,
P.~K., and
  Buchler, B.~C.
\newblock Coherent optical pulse sequencer for quantum applications.
\newblock {\em Nature}{ \bf 461}, 241 (2009).

\bibitem{Moiseev2001}
Moiseev, S.~A. and Kr\"oll, S.
\newblock Complete reconstruction of the quantum state of a single-photon wave
  packet absorbed by a {D}oppler-broadened transition.
\newblock {\em Phys. Rev. Lett.}{ \bf 87}, 173601 (2001).

\bibitem{Kraus2006}
Kraus, B., Tittel, W., Gisin, N., Nilsson, M., Kr\"{o}ll, S., and
Cirac, J.~I.
\newblock Quantum memory for nonstationary light fields based on controlled
  reversible inhomogeneous broadening.
\newblock {\em Phys. Rev. A}{ \bf 73}, 020302 (2006).

\bibitem{Hetet2008}
H\'etet, G., Longdell, J.~J., Alexander, A.~L., Lam, P.~K., and Sellars, M.~J.
\newblock Electro-optic quantum memory for light using two-level atoms.
\newblock {\em Phys. Rev. Lett.}{ \bf 100}, 023601 (2008).

\bibitem{Afzelius2009a}
Afzelius, M., Simon, C., de~Riedmatten, H., and Gisin, N.
\newblock Multimode quantum memory based on atomic frequency combs.
\newblock {\em Phys. Rev. A}{ \bf 79}, 052329 (2009).

\bibitem{Riedmatten2008}
de~Riedmatten, H., Afzelius, M., Staudt, M.~U., Simon, C., and Gisin, N.
\newblock A solid-state light-matter interface at the single-photon level.
\newblock {\em Nature}{ \bf 456}, 773 (2008).

\bibitem{Simon2007a}
Simon, J., Tanji, H., Thompson, J.~K., and Vuletic, V.
\newblock Interfacing collective atomic excitations and single photons.
\newblock {\em Phys. Rev. Lett.}{ \bf 98}, 183601 (2007).

\bibitem{Zhao2009}
Zhao, B., Chen, Y.-A., Bao, X.-H., Strassel, T., Chuu, C.-S., Jin, X.-M.,
  Schmiedmayer, J., Yuan, Z.-S., Chen, S., and Pan, J.-W.
\newblock A millisecond quantum memory for scalable quantum networks.
\newblock {\em Nat Phys}{ \bf 5}, 95 (2009).

\bibitem{Zhao2009a}
Zhao, R., Dudin, Y.~O., Jenkins, S.~D., Campbell, C.~J., Matsukevich, D.~N.,
  Kennedy, T. A.~B., and Kuzmich, A.
\newblock Long-lived quantum memory.
\newblock {\em Nat Phys}{ \bf 5}, 100 (2009).

\bibitem{Lan2009}
Lan, S.-Y., Radnaev, A.~G., Collins, O.~A., Matsukevich, D.~N., Kennedy, T.~A.,
  and Kuzmich, A.
\newblock A multiplexed quantum memory.
\newblock {\em Opt. Express}{ \bf 17}, 13639 (2009).

\bibitem{Nunn2008}
Nunn, J., Reim, K., Lee, K.~C., Lorenz, V.~O., Sussman, B.~J., Walmsley, I.~A.,
  and Jaksch, D.
\newblock Multimode memories in atomic ensembles.
\newblock {\em Phys. Rev. Lett.}{ \bf 101}, 260502 (2008).

\bibitem{chaneliere-2009}
Chaneli\`{e}re, T., Ruggiero, J., Bonarota, M., Afzelius, M., and
Le Gou\"{e}t, J.~L.
\newblock Efficient light storage in a crystal using an atomic frequency comb.
\newblock {\em arXiv:0902.2048}{ \bf } (2009).

\bibitem{Afzelius2009b}
Afzelius, M., Usmani, I., Amari, A., Lauritzen, B., Walther, A., Simon, C.,
  Sangouard, N., Min\'{a}\v{r}, J., de~Riedmatten, H., Gisin, N., and Kr\"oll, S.
\newblock Demonstration of atomic frequency comb memory for light with
  spin-wave storage.
\newblock {\em Phys. Rev. Lett }{ \bf 104 }, 040503 (2010).

\bibitem{Amari2009a}
Amari, A., Walther, A., Sabooni, M., Huang, M., Kr\"oll, S., Afzelius, M.,
  Usmani, I., Lauritzen, B., Sangouard, N., de~Riedmatten, H., and Gisin, N.
\newblock Towards an efficient atomic frequency comb quantum memory.
\newblock {\em arXiv:0911.2145}{ \bf } (2009).

\bibitem{Hesselink1979}
Hesselink, W.~H. and Wiersma, D.~A.
\newblock Picosecond photon echoes stimulated from an accumulated grating.
\newblock {\em Phys. Rev. Lett.}{ \bf 43}, 1991 (1979).

\bibitem{Carlson1984}
Carlson, N. W., Bai, Y. S., Babbitt, W. R., and Mossberg, T. W.
\newblock Temporally programmed free-induction decay
\newblock \emph{Phys. Rev. A} \textbf{30}, 1572 (1984).

\bibitem{Mitsunaga1991}
Mitsunaga, M., Yano, R., and Uesugi, N.
\newblock Spectrally programmed stimulated photon echo.
\newblock {\em Opt. Lett.}{ \bf 16}, 264 (1991).

\bibitem{Schwoerer1994}
Schwoerer, H., Erni, D., Rebane, A., and Wild, U. P.
\newblock Subpicosecond pulse shaping via spectral hole-burning.
\newblock {\em Opt. Commun.}{ \bf 107}, 123 (1994).

\bibitem{Merkel1996}
Merkel, K. D. and Babbitt, W. R.
\newblock Optical coherent-transient true-time-delay regenerator.
\newblock {\em Opt. Lett.}{ \bf 21}, 1102 (1996).

\bibitem{Tian2001}
Tian, M., Reibel R., and Babbitt, W. R.
\newblock Demonstration of optical coherent transient true-time delay at 4 Gbits/s.
\newblock {\em Opt. Lett.}{ \bf 26}, 1143 (2001).

\bibitem{Macfarlane1998}
Macfarlane, R. M., Meltzer, R. S., Malkin B. Z.
\newblock Optical measurement of the isotope shifts and hyperfine and superhyperfine interactions of Nd in the solid state.
\newblock {\em Phys. Rev. B}{ \bf 58}, 5692 (1998).

\bibitem{Bertaina2007}
Bertaina, S., Gambarelli, S., Tkachuk, A., Kurkin, I. N., Malkin, B., Stepanov, A., Barbara, B.
\newblock Rare-earth solid-state qubits.
\newblock {\em Nature Nanotechnology}{ \bf 2}, 39 (2007).

\bibitem{Lauritzen2009}
Lauritzen, B., Min\'{a}\v{r}, J., de Riedmatten, H., Afzelius, M., Sangouard, N., Simon, C., and Gisin, N.
Telecommunication-wavelength solid-state memory at the single photon level.
\newblock
\newblock {\em arXiv:0908.2348}{ \bf } (2009).

\end{thebibliography}

\begin{thebibliography}{10}

\bibitem{deRiedmatten2008}
de~Riedmatten, H., Afzelius, M., Staudt, M.~U., Simon, C., and Gisin, N.
\newblock A solid-state light-matter interface at the single-photon level.
\newblock {\em Nature}{ \bf 456}(7223), 773--777 December  (2008).
\bibitem{Afzelius2009}
Afzelius, M., Simon, C., de~Riedmatten, H., and Gisin, N.
\newblock Multimode quantum memory based on atomic frequency combs.
\newblock {\em Phys. Rev. A}{ \bf 79}(5), 052329--9 May  (2009).

\end{thebibliography}

\end{document}